\documentclass[]{aastex631}
\usepackage{graphicx} % Required for inserting images
\usepackage{amsmath}

\begin{document}
\title{Optimal mitigation of random telegraph noise for improved photometry at high frame rates}

\author[0000-0002-7191-4403]{Christopher Layden}
\altaffiliation{Corresponding author;  \href{mailto:clayden7@mit.edu}{clayden7@mit.edu}}
\affiliation{MIT Kavli Institute for Astrophysics and Space Research, Massachusetts Institute of Technology, 77 Massachusetts Ave, Cambridge, MA 02139, USA}
\affiliation{MIT Department of Physics, 77 Massachusetts Ave., Cambridge, MA 02139, USA}

\author[0000-0002-8590-007X]{Daniel-Rolf Harbeck}
\affiliation{Las Cumbres Observatory, Goleta, CA, USA}

\author[0009-0008-7670-3630]{Tejus Deo-Dixit}
\affiliation{MIT Department of Physics, 77 Massachusetts Ave., Cambridge, MA 02139, USA}

\author[0000-0002-4585-9981]{Nathan Lourie}
\affiliation{MIT Kavli Institute for Astrophysics and Space Research, Massachusetts Institute of Technology, 77 Massachusetts Ave, Cambridge, MA 02139, USA}

\author[0000-0001-8467-9767]{Gabor Furesz}
\affiliation{Astrophysics \& Space Center, Schmidt Sciences, New York, NY 10011, USA}
\affiliation{MIT Kavli Institute for Astrophysics and Space Research, Massachusetts Institute of Technology, 77 Massachusetts Ave, Cambridge, MA 02139, USA}

\author[0000-0002-7226-836X]{Kevin Burdge}
\affiliation{MIT Kavli Institute for Astrophysics and Space Research, Massachusetts Institute of Technology, 77 Massachusetts Ave, Cambridge, MA 02139, USA}
\affiliation{MIT Department of Physics, 77 Massachusetts Ave., Cambridge, MA 02139, USA}

\begin{abstract}
    Random telegraph noise (RTN) is a major contributor to read noise in many CMOS image sensors considered for astronomical use. While scientific CMOS image sensors deliver lower read noise than traditional charge-coupled devices, mitigating RTN would widen this gap and enable more precise photometry when using the fast readout rates achievable by CMOS image sensors. We report the levels of RTN in three CMOS image sensors used in astronomical instruments: the Sony IMX455, Gpixel GSENSE400, and Fairchild Imaging HWK4123. For the IMX455 in a high gain mode, RTN is the dominant source of pixels with high read noise and increases the overall read noise floor by $\gtrsim 20\%$. RTN is present in the GSENSE400 and HWK4123 but to smaller effects. We compare two strategies for RTN mitigation: masking pixels exhibiting RTN or using a new algorithm for correcting RTN jumps. For faint ($< 3\,\mathrm{e}^-$/pix/frame) observations of a stellar field with the IMX455, both masking and our algorithm improved the signal-to-noise ratio (SNR) of light curves by $> 5\%$ on average. Larger improvements were achieved for sources falling on multiple RTN pixels. Our algorithm outperforms masking, especially when the point spread function is undersampled, masked pixels are near the source center, or read noise and shot noise are comparable. In such cases, masking may even deteriorate photometric precision. In other cases, masking remains an effective RTN mitigation technique. We have made available our software for identifying RTN pixels, parametrizing their bias level distributions, and applying our correction algorithm.
\end{abstract}


\section{Introduction}
Many recently commissioned or proposed instruments for ultraviolet (UV), optical, and infrared (IR) astronomy employ complementary metal-oxide-semiconductor (CMOS) image sensors. Custom UV-sensitive CMOS sensors have been developed for the upcoming Ultraviolet Transient Astronomy Satellite (ULTRASAT) and Ultraviolet Explorer (UVEX) missions \citep{Bastian_Querner_2021,greffe:2022}. Commercially available CMOS image sensors have increasingly been selected for ground-based and space-based optical instruments, including proto-Lightspeed on the Magellan Clay Telescope \citep{layden:2026}, the Large Array Survey Telescope \citep[LAST;][]{Ofek_2023}, the small telescope fleet of Las Cumbres Observatory \citep[LCO;][]{brown2013, harbeck:2024}, and the Lazuli Space Observatory \citep{Roy2026}. Hybridized CMOS image sensors (notably the HxRG series outlined in \citet{blank:2011}) have become near-ubiquitous for infrared astronomy, finding use in some of today's premier space telescopes, such as the James Webb Space Telescope \citep[JWST;][]{Rauscher_2007} and the Roman Space Telescope \citep{mosby:2020}.

Random telegraph noise (RTN; sometimes called ``salt-and-pepper" or popcorn noise) has been observed in all of these types of CMOS image sensors and, specifically, in the detectors of JWST, LAST, the LCO small telescope fleet, and the Lazuli Space Observatory \citep{greffe:2022,ozdogru:2025,bevidas:2025}. If this noise component were removed, then the read noise floor of these instruments could be significantly improved, allowing higher signal-to-noise ratios (SNR) when observing faint scenes or using high frame rates. RTN mitigation is especially salient for space telescopes, as radiation damage is known to increase the prevalence of RTN \citep{hopkins:1993,liu:2023,antonsanti:2022}.

% Moreover, such mitigation could improve the viability of HxRG detectors that were not selected for use due to elevated levels of RTN \citep{antonsanti:2024}, potentially reducing the cost of infrared astronomy. ZZZ this mitigation isn't yet super applicable to HxRG because they mostly use up-the-ramp sampling instead of CDS, and read noise swamps a lot of RTN jumps

RTN arises in solid-state transistors from the capture or release of electrons by carrier traps near the Si/SiO$_2$ interface \citep{wang:2006}. In a charge-coupled device (CCD), the source follower transistor of the output node(s)---where photogenerated charge is converted to a readable voltage---may contain many such traps. The superposition of these traps changing occupancy can result in $1/f$ noise \citep{vanderziel:1970}, which can dominate the read noise of CCDs \citep{hegyi:1980}. As device fabrication has improved, allowing for ever-smaller transistors with ever-fewer defects, the number of interface traps in each transistor has fallen. This is true in particular for CMOS image sensors, in which every pixel has its own source follower transistor, typically with size of $\lesssim 1\,\mathrm{\mu m}$. In scientific CMOS image sensors, each source follower transistor may have zero to a few traps. When only $O(1)$ traps are present, the bias level of a source follower transistor will bounce between a set of discrete values corresponding to the possible occupation states \citep{janesick_i:2006}.

In most imaging applications, the signal in each pixel is sampled twice: once before photogenerated charge has been transferred to the gate of the source follower transistor and once after charge transfer. The first reading is subtracted from the second to recover the signal level. In this process, called correlated double sampling (CDS), much excess read noise is removed, and only changes in trap occupation that occur between the samples will affect the signal value \citep{wang:2006}. However, such changes in occupation between samples can still contribute significantly to an image sensor's read noise, as RTN.

\begin{figure}
    \centering
    \includegraphics[width=0.95\linewidth]{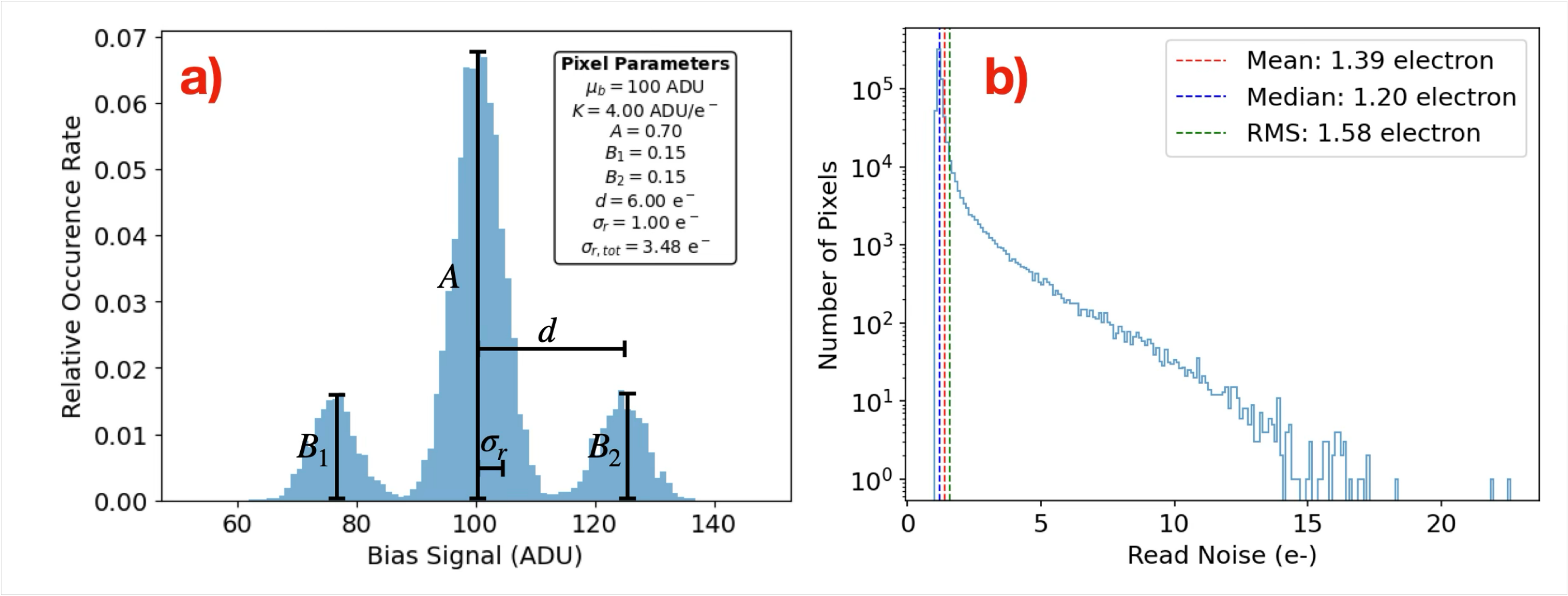}
    \caption{\textbf{a)} Histogram of bias levels reported by a simulated image sensor pixel with random telegraph noise. As is typical in image sensors to avoid clipping negative values, a mean bias level of $\mu_b=100\,$ADU is added. The gain of the simulated pixel is set to 10~ADU/e$^-$. The inherent read noise of the pixel $\sigma_r$ is broadened by the RTN to a larger value $\sigma_{r,tot}$. \textbf{b)} Read noise measured in 10,000 pixels of an IMX455 sensor. The long tail, largely consisting of pixels with RTN, significantly increases the RMS read noise.}
    \label{fig:rtn_hist}
\end{figure}

The expected bias level distribution of a pixel with one RTN-generating trap is a sum of three Gaussians, as depicted in Fig.~\ref{fig:rtn_hist}a. The central Gaussian, with relative height $A$, represents bias level readings where trap occupation did not change between samples; its standard deviation $\sigma_r$ represents the read noise the pixel would have if the trap was not present. The Gaussians to the left and right represent measurements where the trap has released or captured a carrier, respectively. These Gaussians have relative heights $B_1$ and $B_2=1-A-B_1$, are equidistant (with spacing $d$) from the central Gaussian, and also have standard deviation $\sigma_r$ (release or capture could conceivably affect the standard deviation, but this slight effect is not often observed). Depending on the timescales of trap transition and pixel readout, $B_1$ can be near zero \cite{wang:2006}. The parameters governing the bias level distribution ($A,B_1,\sigma_r,d$ and others if more than one trap is present) can be different for all of the pixels with RTN in a given sensor, depending on the location of the trap in each transistor and the local distribution of dopant atoms \citep{lee:2003,janesick_i:2006}. RTN is a significant noise contributor in instances where the spacing $d$ is larger than both $\sigma_r$ and the Poissonian fluctuations in the level of photogenerated charge ($\sim \sqrt{\lambda}$ for a mean photoelectron number $\lambda$).

% {\bf XXX DRH: This article focuses on pixels where $B_1 \simeq B_2$.}
% Note from Chris: two of the sensors analyzed actually have a large fraction of near-zero B1

\citet{ozdogru:2025} and \citet{bevidas:2025} have developed methods for identifying RTN in pixels of CMOS and hybrid CMOS image sensors. Across such sensors, RTN was identified in $\approx$~1--5\% of pixels. However, the total contribution that RTN makes to the noise budget of CMOS image sensors has not been reported. Because $d$ is often much larger than $\sigma_r$, the overall read noise of a pixel with RTN, $\sigma_{r,tot}$, can be significantly larger than the read noise of pixels without RTN. For this reason, the distribution of pixel read noise values in many scientific CMOS image sensors has a long tail composed of pixels demonstrating RTN of varying severity. This tail can significantly degrade the root mean square (RMS) read noise of a sensor while slightly degrading the mean read noise and negligibly affecting the median read noise. This is demonstrated in Figure~\ref{fig:rtn_hist}b, which shows the pixel read noise distribution for a Sony IMX455 image sensor. As the total read noise for aperture photometry is found by summing in quadrature the read noise from individual pixels, RMS read noise is the relevant figure of merit for assessing sensor noise performance. Users of CMOS image sensors should therefore be wary of advertised median and mean read noise values. 

As an interim step to quantify the effects of RTN, the data processing pipeline at Las Cumbres Observatory seeds the read noise not only with a global read noise value (as is common for CCD data reduction) but with a map that reflects the read noise $\sigma_{r,tot}$ in each individual pixel \citep{McCully2018,harbeck:2024}. However, even this may underestimate the effects that RTN jumps can have on photometry when only a few exposures are taken, because $\sigma_{r,tot}$ underestimates the peak-to-peak separation of values reported by RTN pixels by $\sim \sqrt{2}$. 

CMOS image sensors offer inherent advantages in readout speed, power consumption, relaxed cooling requirements, and radiation tolerance relative to CCDs \citep[$T\gtrsim -25^\circ$C;][]{Alarcon_2023,Khandelwal_2024,layden:2025}. Inexpensive, commercially available CMOS image sensors can also deliver RMS read noise below $2\,\mathrm{e^-}$ \citep{Khandelwal_2024}, significantly better than conventional CCDs. If the excess noise from RTN pixels in CMOS image sensors can be mitigated, either in image processing or by advances in fabrication, then such sensors would strengthen this advantage compared to CCDs.

% {\bf XXX DRH: Note that a rms for a pixel's readnoise in the presence of RTN would underestimate peak to peak values by a factor of ~ sqrt(2). x}

% {\bf XXX DRH: As an interim step, the data processing pipeline at Las Cumbres Observatory therefore seeds the readnoise not only with a global readnoise value as it is common for CCD data reduction, but with a noise map that reflects each individual CMOS pixels rms noise \citep{McCully2018,harbeck:2024}.} 
% 

Methods for mitigating the effects of RTN have been proposed, including applying a median filter, which cosmetically improves salt-and-pepper features in images but correlates nearby pixels and throws away pixel data, degrading photometric precision. In another approach, referred to here as masking, all RTN pixels are treated as defective and excluded from analysis. The most sophisticated RTN mitigation strategy for astronomy to date, proposed by \citet{Alarcon_2023}, involves subtracting a median-filtered frame from each image and identifying pixels with residuals above a fixed value as having random telegraph jumps. Identified jumps could be masked or replaced in some user-defined way. This strategy has a few downsides. First, it treats all pixels equally, despite the diversity in bias level distributions demonstrated by RTN pixels. Without knowledge of individual pixel bias level distributions, using a fixed cutoff might not reliably or accurately identify RTN jumps, and it is impossible to know by how much a jump from a given pixel should be corrected. Second, the strategy is agnostic to the photon flux. While it may be effective for near-zero flux levels, whenever the photon shot noise grows comparable to the fixed cutoff value, it cannot be used reliably. Finally, median filtering may not be robust for critically sampled or undersampled point-spread functions (PSFs).

An ideal RTN correction strategy would consider the bias level distribution of every pixel to reliably identify and correct RTN jumps. It would also adapt given the average photon flux to ensure that photon fluctuations are not misidentified as jumps. Here, we propose such a strategy and for the first time demonstrate concretely how photometric precision can be improved by the mitigation of RTN. We also compare how this algorithm performs relative to the much simpler masking approach.

First, we introduce a few CMOS image sensors that have been considered or are in use for astronomical applications, and we describe our methods for identifying and parameterizing RTN in these sensors. We then detail our algorithm for identifying and correcting instances of random telegraph jumps in astronomical images. Next, we report the level of RTN measured in each sensor and the expected photometric improvement to be gained by our RTN correction algorithm and by pixel masking. We compare these expected improvements to the improvements seen after applying the correction strategies to observations of a stellar field taken with Sony IMX455 cameras mounted on the small telescope network of LCO. Finally, we note how astronomers can access and use our RTN correction algorithm to potentially gain an immediate boost in photometric precision for instruments using detectors with RTN.

\section{Methods}
\subsection{Correction algorithm development}
When the bias level distribution is known for all pixels with RTN, it becomes possible to identify random telegraph jumps and appropriately correct these jumps. We developed an algorithm to perform these steps, using statistical considerations to ensure the reliability and efficacy of random telegraph jump identification. In Sec.~\ref{sec:algorithm}, we describe this algorithm in more detail.

\subsection{Sensors under test}
In this work, we focus on commercial off-the-shelf (COTS) cameras housing CMOS image sensors that have been widely used or considered for optical imaging applications in astronomy. Namely, we study the Sony IMX455, the Gpixel GSENSE400, and the Fairchild Imaging HWK4123 image sensors.

The IMX455 image sensor is a high resolution ($9600\times 6422$ pixels) CMOS sensor with 3.76~µm pixels and is employed in many commercially available camera packages, including the Atik APX60, the ZWO ASI6200, and the QHYCCD QHY600M. With its relatively large size, full frame rates above 2.5~Hz (for the QHY600M, limited by USB3 data rates), peak quantum efficiency (QE) above 80\%, low dark current, and RMS read noise as low as 1.42~e$^-$ when using a high gain setting \citep{Alarcon_2023}, this sensor has been selected for or integrated in serious astronomical instruments, including LAST \citep{Ofek_2023}, the small telescope fleet of LCO \citep{harbeck:2024}, the CubeSats for Rapid Infrared and Optical Surveys Mission \citep[CuRIOS;][]{Gulick_2024}, and the Lazuli Space Observatory \citep{Roy2026}. The IMX455 and other sensors in the Sony IMX series have been observed to have significant levels of RTN \citep{Alarcon_2023,harbeck:2024,ozdogru:2025}. For this work, we used three IMX455 sensors housed in QHY600M cameras that are mounted in the LCO small telescope fleet. We will refer to these cameras as QHY600M-A, QHY600M-B, and QHY600M-C. QHY600M-A and QHY600M-B are mounted at McDonald Observatory, Texas, while QHY600M-C is mounted at Haleakal\=a Observatory, Hawaii. We also conducted on-sky observations with these cameras, as discussed in Sec.~\ref{sec:testing_methods}.

The GSENSE400 image sensor has $2048 \times 2048$ pixels of size 11~µm and is also available in multiple commercially available cameras, including the Andor Marana and the QHYCCD QHY42. These cameras deliver full frame rates up to 48~Hz, peak quantum efficiency above 90\%, and RMS read noise of $\sim 1.7\,\mathrm{e}^-$ when using a high gain mode, but with non-negligible dark current \citep{Khandelwal_2024,qiu:2021,apergis:2025}. With its pixel size closely matching astronomical CCDs, this sensor been considered as a potential upgrade to such detectors \citep{apergis:2025,qiu:2021}.

The HWK4123 sensor has $4096\times 2304$ pixels of size 4.6~µm and is commercially available in the Hamamatsu ORCA-Quest~2 camera. It is one of the only widely available CMOS image sensor designed to deliver deep sub-electron read noise (DSERN), with its ultra-low capacitance floating diffusion nodes providing ultra-high charge-to-voltage conversion gain \citep{gallagher:2024,cho:2023}. This camera delivers frame rates up to 120~Hz, low dark current, and an RMS read noise of $0.3\,\mathrm{e}^-$ when using the ultra-quiet readout mode \citep{Khandelwal_2024,gallagher:2024,layden:2026}. As the first available CMOS image sensor potentially capable of resolving individual photons, this camera has been integrated into the VAMPIRES instrument \citep{lucas:2024}, the ORCA-TWIN instrument \citep{roth:2025}, and the proto-Lightspeed instrument \citep{layden:2026}. It too has been proposed for use in the Lazuli Space Observatory \citep{Roy2026}.

\subsection{Camera settings}
\label{sec:camera_settings}
A sensor's RTN properties may depend strongly on the readout mode and the temperature \citep{wang:2006,bevidas:2025}, so these variables should remain fixed during RTN measurement and correction. For the three QHY600M cameras that we tested, we used the 16-bit ``High Gain Mode" (readout mode 1) with gain setting 60 to minimize the read noise (specified as $\sim1.69\,\mathrm{e}^-$ in this mode) while maintaining adequate full well capacity (specified as $\sim20,000\,\mathrm{e}^-$ in this mode). We maintained these cameras at a temperature of $2^\circ$C when in use to prevent ice formation in the detector chamber while still yielding negligible dark current for short exposures. For the QHY42 camera, we used the 12-bit standard readout mode with a gain setting of 7, an offset of 1000\,ADU, and a temperature of $2^\circ$C. For the Hamamatsu ORCA-Quest~2 camera, we used the 16-bit ultra-quiet scan, area readout mode at a fixed temperature of $-20^\circ$C. For all cameras under test, we recorded and analyzed data from only the central $800\times 800$\,pixel region to save time in data processing and analysis.

\subsection{Read noise and gain measurement}
\label{sec:methods_read_noise}
To study the read noise delivered by the cameras under test and allow for later identification of RTN, we collected 5,000 bias frames (i.e., frames taken at minimum exposure time and with the lens cap on). The minimum exposure times for the QHY600M, QHY42, and ORCA-Quest~2 cameras were 1~ms, 0.05~ms, and 34~µs, respectively. Using these stacks of bias frames, we calculated the standard deviation in the bias level (i.e., the read noise) of each individual pixel, in analog-to-digital units (ADU). We calculated the median, mean, and RMS read noise of the cameras using the distributions of read noise across the pixels.

To convert these measurements to more meaningful units of e$^-$, we used the photon transfer method to derive the conversion gain $K$ of each sensor in ADU/e$^-$ \citep{emva1288}. In this method, we uniformly illuminated each camera and collect images with increasing exposure times, taking at least ten images at each exposure time. For the QHY42 and ORCA-Quest~2 cameras, we use the uniform light source presented in \cite{layden:2025}. We uniformly illuminate QHY600M-A using dome flats. Using these data, we calculate the mean and variance in signal reported by each pixel at each exposure time, then calculate the averages of these values across the sensor. We then plot the variance vs. mean value at each exposure time and perform a linear fit. The slope of this fit gives the conversion gain, in ADU/e$^-$. We report the read noise distribution and gain for the sensors under test in Sec.~\ref{sec:results_read_noise}.

\subsection{RTN pixel identification and parametrization}
\label{sec:identification}
We then sought to identify pixels with RTN and parametrize the bias level distributions of these pixels, using the first 5000 bias frames collected for each sensor and the measured conversion gain. Our procedure is described in more detail in Appendix~\ref{sec:app_fitting} and is based on the methods of \cite{ozdogru:2025}. We applied the Anderson-Darling (AD) normality test to each pixel's bias readings. For each pixel failing this normality test, we calculated a histogram of its bias readings. We attempted to fit a triple-Gaussian distribution to each such histogram, consisting of independent parameters $\mu_b, A, B_1, d,\sigma_r$ as depicted in Fig.~\ref{fig:rtn_hist}. If the triple-Gaussian fit was acceptable (see Appendix~\ref{sec:app_fitting}), the pixel was identified as having correctable RTN. We did not attempt to fit more complicated bias level distributions, like the 5-Gaussian distribution that can arise when 2 traps are present in the pixel, because doing so would further complicate the correction algorithm. For each sensor under test, we saved a Flexible Image Transport System (FITS) file containing the best-fit RTN parameters. This file contains a $6\times M\times N$ array for $M$ rows and $N$ columns of pixels in the tested sensor, with the first 5 frames corresponding to the 5 triple-Gaussian parameters. The sixth frame saved in this file contains a value $\lambda_{max}$ for each pixel, derived from $d$ and $\sigma_r$ and corresponding to the maximum photoelectrons per frame where RTN jumps can still be corrected. This value is defined in Sec.~\ref{sec:jump_id}. Any pixels not identified as having RTN have \texttt{NaN} entries in the saved file.

In Sec.~\ref{sec:rtn_id_results}, we report the fraction of pixels demonstrating RTN in each sensor, in particular checking for consistency across the three QHY600M cameras. We also analyze the spatial distribution of RTN pixels in each sensor, observing whether they demonstrate any pattern.

\subsection{Prediction of RTN mitigation effectiveness}
\label{sec:prediction_methods}
By replacing the read noise of RTN pixels with their fit $\sigma_r$ values, we found the effective RMS read noise of each sensor assuming all RTN jumps could be perfectly corrected in these pixels. This value is an upper limit to the RMS read noise that each sensor would achieve if no RTN was present, as our identification may not catch pixels with more than one RTN-generating trap or with small RTN jumps.

We also estimated the effective RMS read noise that will be achieved after applying our correction algorithm as a function of the average number of photoelectrons per frame $\lambda$. For each pixel, we modeled the effective read noise as varying between two limits: the read noise without RTN $\sigma_r$ (achieved at $\lambda=0$) and the original (uncorrected) read noise $\sigma_{r,tot}$ (achieved at $\lambda = \lambda_{max}$). For intermediate values of $\lambda$, we linearly interpolate between these two limits. When $\lambda > \lambda_{max}$, we assume the correction provides no benefit and use $\sigma_{r,tot}$. At each value of $\lambda$, we report the RMS of the effective read noise distribution across all pixels.

We also conducted Monte Carlo simulations to probe how the correction algorithm and RTN pixel masking, respectively, affect the SNR of photometric observations. We performed these simulations varying the source brightness, PSF size, fraction of pixels with RTN in a sensor, and properties of these RTN pixels. We provide the results of these simulations and recommendations for when to use our correction algorithm or pixel masking in Sec.~\ref{sec:rtn_predicted_effects}.

\subsection{Correction algorithm testing}
\label{sec:testing_methods}
% {\bf XXX DRH: This might better be relocated to Sect 4? CPL: I moved things around, hopefully the flow is better now.}

To confirm the improvements that the RTN correction algorithm and masking can deliver for astronomy, we used astronomical observations conducted with the QHY600M cameras mounted on the LCO small telescope network. In Sec.~\ref{sec:on_sky}, we describe these observations and the benefits gleaned from applying the either RTN correction algorithm or RTN pixel masking.

\section{RTN Correction Algorithm}
\label{sec:algorithm}
The RTN correction algorithm compares the signal reported by a pixel to an estimate of its expected signal level, determines whether the reported signal corresponds to an RTN jump, and corrects such jumps. We must take care to only alter the scientific data when we have high confidence that a jump has occurred: an algorithm that too aggressively ``corrects" the data might appear to improve the SNR of a light curve but may actually wash out astrophysical signals. Here we outline the steps of the algorithm. We assume the user has followed the procedure described in Sec.~\ref{sec:identification} to identify pixels with RTN and parametrize their bias level distributions, resulting in a FITS file containing $\mu_b, A, B_1, d, \sigma_r$ for each pixel, with $d$, and $\sigma_r$ in units of e$^-$. Sec.~\ref{sec:usage} outlines how a user may access and use the software to apply this correction algorithm.

\subsection{Expected signal level determination}
\label{sec:expected_lambda}
It is relatively easy to identify RTN jumps in bias frames, as one may readily compare a pixel's signal level $S$ (in ADU) to its central Gaussian mean $\mu_b$. The situation is harder in illuminated images, where the ``expected" signal level for the central Gaussian is unknown and shot noise broadens all three Gaussian peaks. However, we can estimate the expected signal level in a pixel, denoted $\mu_b +K\lambda$ (with $\lambda$ the expectation value of the number of photoelectrons in the pixel), to compare to $S$. As long as the magnitude of an RTN jump is large relative to shot noise and other noise sources expected near this signal level, the jump can be identified and corrected. This estimate need not be perfect, as it is only used as a reference value to determine whether a correction should be applied, never the magnitude of such a correction. Our algorithm can apply one of two techniques for find a pixel's reference signal level $\mu_b + K\lambda$, suitable for different types of astronomical observations.

In the first technique, the reference value $\mu_b+K\lambda$ for a pixel in frame number $i$ is found as the median of the pixel's values in frames $[i - N/2, i +N/2]$ for window size $N$, excluding frame $i$ (we have typically adopted $N\approx 10$). This technique is suitable for critically- or under-sampled sources and when sequences of frames much larger than $N$ are taken, as the first and last $N/2$ frames are not corrected. It is suitable for fast imaging of rapidly varying sources, as long as such variability is either well sampled across $\approx N$ frames or is small in magnitude in each pixel relative to shot noise. As a precaution to avoid misidentification of variability as RTN jumps, if this technique is used, the standard deviation of the values in frames $[i - N/2, i +N/2]$ (excluding frame $i$) is calculated. If this standard deviation is larger than $1.5\times K\sqrt{\sigma_{r,tot}^2+\lambda}$, then no correction is attempted. This cutoff means that for a given pixel to be corrected, the mean source brightness should be varying by $\lesssim \sqrt{\lambda}$ over $N$ frames.

The second technique involves applying a spatial median filter, as suggested by \citet{Alarcon_2023}. Here, a reference frame is generated by replacing every pixel's signal level with the median of its neighboring pixels. This technique does not rely on any frames preceding or succeeding the frame of interest, so it is useful for sources with any degree of variability. However, for a spatial median to be effective, pixels should over-sample the PSF. Otherwise, around sources, the median filter will give an inaccurate estimate of the expected signal level in each pixel.

By applying one of these technique, a reference value is generated for each RTN pixel, for each frame in an image stack. By subtracting $\mu_b$ and dividing by $K$, this reference value is expressed in photoelectrons $\lambda$.

\subsection{Jump identification and correction}
\label{sec:jump_id}
The algorithm then determines whether a jump occurred in an RTN pixel, given the pixel's reference value $\lambda$; its bias level distribution parameters $\mu_b$, $\sigma_r$, $d$; and its actual signal value expressed in e$^-$, $s=(S-\mu_b)/K$. To make this determination, we consider the cumulative density function (CDF) $F(x)$ of the Poisson-Gaussian convolution with mean value $\lambda$ and Gaussian spread $\sigma_r$: 

% To determine whether a measured signal $S$ is consistent with coming from one of the side Gaussians of a pixel's bias level distribution, we consider the cumulative density function (CDF) $F(s)$ of the Poisson-Gaussian convolution with mean value $\lambda$ and Gaussian spread $\sigma_r$: 
\begin{equation}
    F(x) = \sum_{k=0}^{\infty} \frac{\lambda^k e^{-\lambda}}{k!} \, \Phi\left(\frac{x - k}{\sigma_r}\right)
    \label{eq:cdf}
\end{equation}
where $\Phi$ is the standard Gaussian CDF. This is the distribution for the signal expected if no jump has occurred. We find the signal levels $s_{low}$ and $s_{high}$ such that $F(s_{low})=\alpha /2$ and $F(s_{high})=1-\alpha/2$ for a chosen significance level $\alpha$.

The pixel is determined to have a negative RTN jump if its signal level $s$ satisfies
\begin{equation}
    s_{low}-d < s < \min{(s_{high} - d, s_{low})}
    \label{eq:low_inequality}
\end{equation}
and a positive RTN jump if
\begin{equation}
    \max{(s_{low} + d, s_{high})} < s <s_{high}+d.
    \label{eq:high_inequality}
\end{equation}
That is, to be corrected, a signal must correspond to the high or low Gaussian side peak at the significance level $\alpha$ but must not correspond to the central peak at the same significance level. For computational efficiency, we created a look-up table that is interpolated at given values $\lambda,\sigma_r$ to return $s_{low}$ and $s_{high}$. We terminated the infinite sum in Eq.~\ref{eq:cdf} at $k=\lambda+10\sqrt{\lambda}+20$. By default, we set $\alpha =0.003$ to robustly prevent false positives.

% Any cosmic ray events, typically causing jumps significantly larger than RTN jumps, likely will not be corrected. A separate cosmic ray rejection technique should therefore be employed if desired.

\begin{figure}
    \centering
    \includegraphics[width=0.99\linewidth]{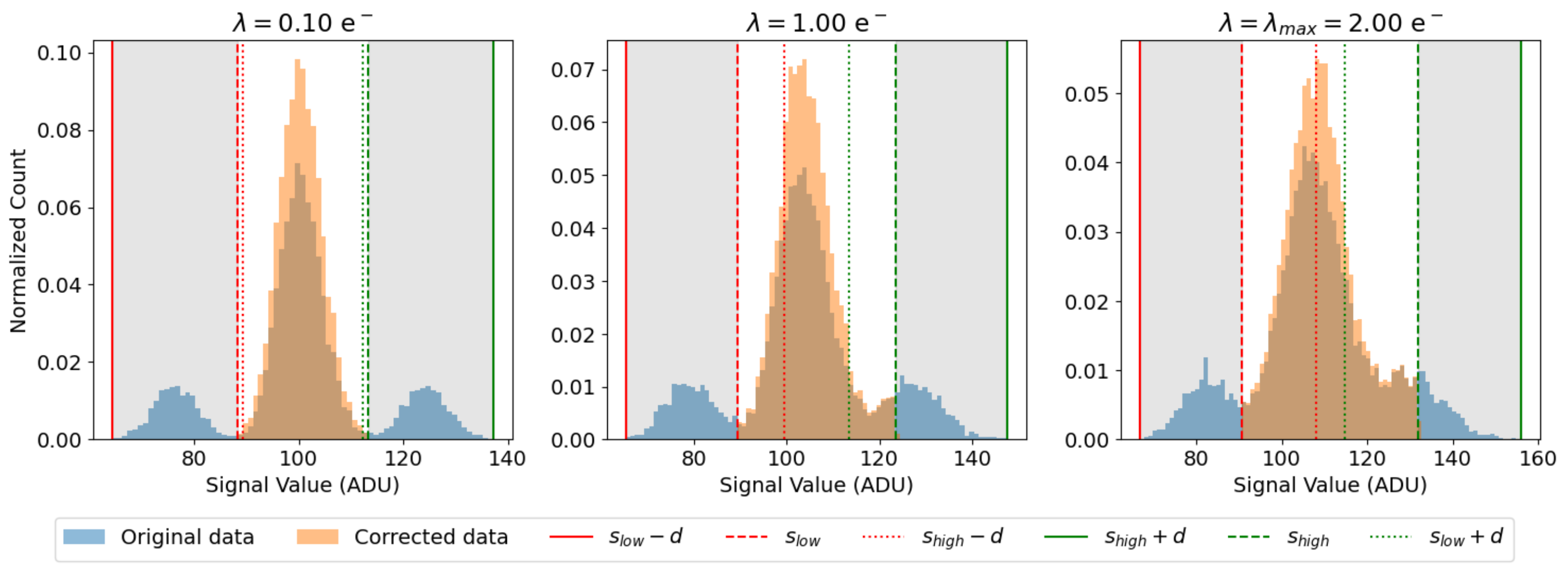}
    \caption{Examples of the correction algorithm applied to data from a simulated pixel at increasing levels of average illumination. The simulated pixel has $\mu_b=100$~ADU, $K=4\,\mathrm{ADU/e}^-,\,A=0.7,\,B_1=0.15,\,d=6\,\mathrm{e}^-,\,\sigma_r=1\,\mathrm{e}^-$. The three plots show original (blue) and corrected (orange) data points for when the pixel collects an average of 0.1~e$^-$, 1~e$^-$, and 2~e$^-$ per frame, respectively. The gray shaded regions where data are corrected are defined by Eq.~\ref{eq:low_inequality} (for low RTN jumps) and Eq.~\ref{eq:high_inequality} (for high RTN jumps). The cutoff values used in those equations are shown by solid and dashed lines (after scaling by $K$ and adding $\mu_b$).}
    \label{fig:algorithm_example}
\end{figure}

If a negative RTN jump is identified, then the signal level $S$ is replaced with $S + Kd$; for a positive jump, by $S-Kd$. The software thereby returns a stack of corrected FITS images. Figure~\ref{fig:algorithm_example} demonstrates the operation of the correction algorithm when applied to simulated observations at increasing illumination levels $\lambda$. The simulated data assume a pixel with the same parameters as in Fig.~\ref{fig:rtn_hist} ($\mu_b=100$~ADU, $K=4\,\mathrm{ADU/e}^-,\,A=0.7,\,B_1=0.15,\,d=6\,\mathrm{e}^-,\,\sigma_r=1\,\mathrm{e}^-$).

As a further constraint to prevent false adjustment of astrophysical signals, no positive RTN jump is identified if if $s_{high}\geq \lambda +d$, and no negative RTN jump is identified if $s_{low}\leq \lambda - d$. That is, when the central Gaussian begins to merge with the left or right peak at significance level $\alpha$, we consider that peak no longer resolvable. We refer to the value $s_{high} - d$ as $\lambda_{max}$ for each pixel, giving the maximum number of photoelectrons per frame where positive RTN jumps may still be corrected. The third panel in Fig.~\ref{fig:algorithm_example} shows how the central and right peaks begin to smear together at $\lambda_{max}$, which for the simulated pixel parameters $\sigma_r=1\,\mathrm{e}^-, d=6\,\mathrm{e}^-$ occurs at $2.0\,\mathrm{e}^-$.

\subsection{Absolute photometry subtlety}
\label{sec:subtlety}
Using the CDF in this manner allows us to reliably separate true and false RTN events. In particular, it accounts for the inherent right-skew of a Poissonian distribution, resulting in $\lambda - s_{low} < s_{high}-\lambda$. If a fixed deviation from $\lambda$ were used to identify RTN events, then either positive flux fluctuations would be more frequently misidentified as RTN jumps or negative RTN jumps would be more frequently missed. However, because negative RTN jumps are more easily corrected than positive jumps, the algorithm may in the long run artificially increase the average value in a pixel, potentially degrading absolute photometric accuracy. To prevent this, the algorithm records how many positive and negative RTN jumps are corrected. Based on the difference between these numbers, it saves a bias correction value for each RTN pixel, which should be subtracted in addition to standard bias subtraction. This bias correction value should not be subtracted from pixels where other processing techniques are applied that could filter out residual RTN jumps, such as sigma-clipping or calculating the median flux in a region (e.g., for background subtraction). In our analysis, because we apply median background subtraction to avoid overestimation of the background from nearby sources, we subtract the bias correction values only for pixels in source apertures.

\section{Results and Discussion}
\subsection{Measured gain and read noise}
We measured conversion gains of 3.26~ADU/e$^-$, 1.52~ADU/e$^-$, and 8.9~ADU/e$^-$ for the QHY600M-A, QHY42, and ORCA-Quest~2 cameras, respectively. As the gain measured for QHY600M-A was within 2\% of the manufacturer specified value (3.32~ADU/e$^-$), we adopt our measured value for the other two QHY600M cameras as well.

\begin{figure}[t!]
    \centering
    \includegraphics[width=\linewidth]{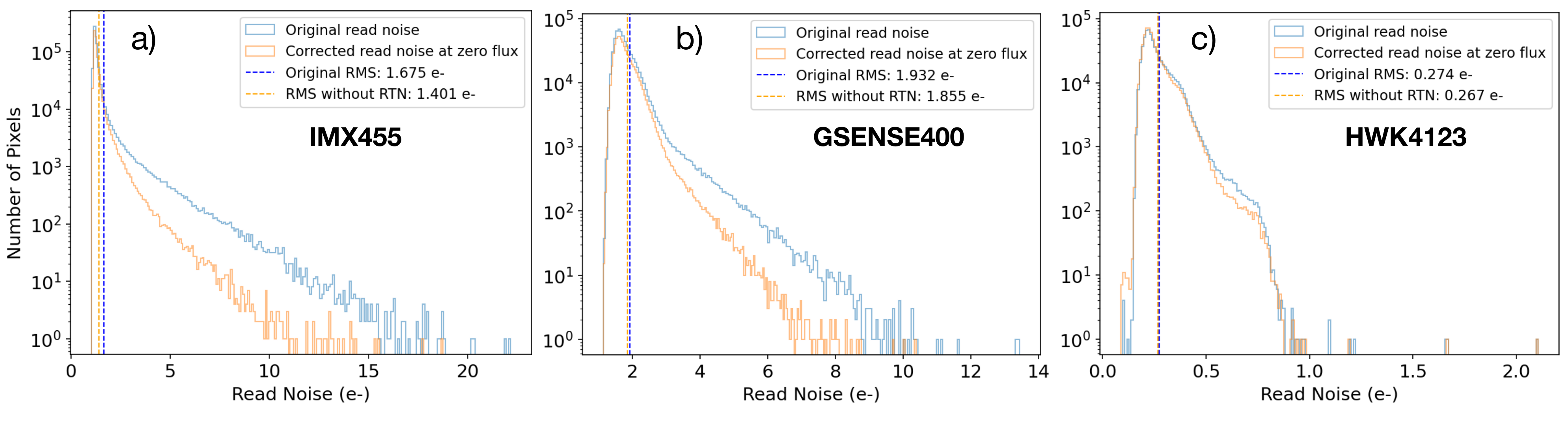}
    \caption{Blue curves show the distributions of read noise in pixels of the cameras under test, \textbf{a)} QHY600M-A, featuring an IMX455 sensor, \textbf{b)} the QHY42, featuring a GSENSE400 sensor, and \textbf{c)} the ORCA-Quest~2, featuring a HWK4123 sensor. For QHY600M-A and the QHY42, a significant number of RTN pixels are identified. Orange curves show the pixel read noise distributions for these cameras assuming all RTN jumps in these pixels can be corrected, which the correction algorithm should nearly achieve when the sensors collect no photoelectrons.}
    \label{fig:read_noise_results}
\end{figure}

The blue curves of Fig.~\ref{fig:read_noise_results} show the distributions of original read noise values in the pixels of QHY600-A, the QHY42 camera, and the ORCA-Quest~2 camera, found using 5000 bias frames. The long tail and difference between the median (1.24~e$^-$) and RMS (1.68~e$^-$) read noise in the QHY600M camera hint at how significantly RTN may be affecting the IMX455 sensor. The read noise distributions for QHY600M-B and QHY600M-C were very similar, with median read noise values of 1.19~e$^-$ and 1.22~e$^-$ and RMS read noise values of 1.57~e$^-$ and 1.61~e$^-$, respectively. The slight differences between these values might be attributable to manufacturing differences, either in pixel quality or in conversion gain. The QHY42 camera also showed an extended (but not as pronounced) read noise distribution tail and a slight difference between the median (1.71~e$^-$) and RMS (1.93~e$^-$) read noise, possibly indicating that RTN is present to a moderate degree. The ORCA-Quest~2 had the least prominent read noise distribution tail and a slight difference between the median (0.24~e$^-$) and RMS (0.27~e$^-$) read noise, indicating that RTN does not significantly affect the HWK4123 sensor.

\label{sec:results_read_noise}

\subsection{RTN identification and parametrization}
\label{sec:rtn_id_results}
\begin{figure}
    \centering
    \includegraphics[width=0.99\linewidth]{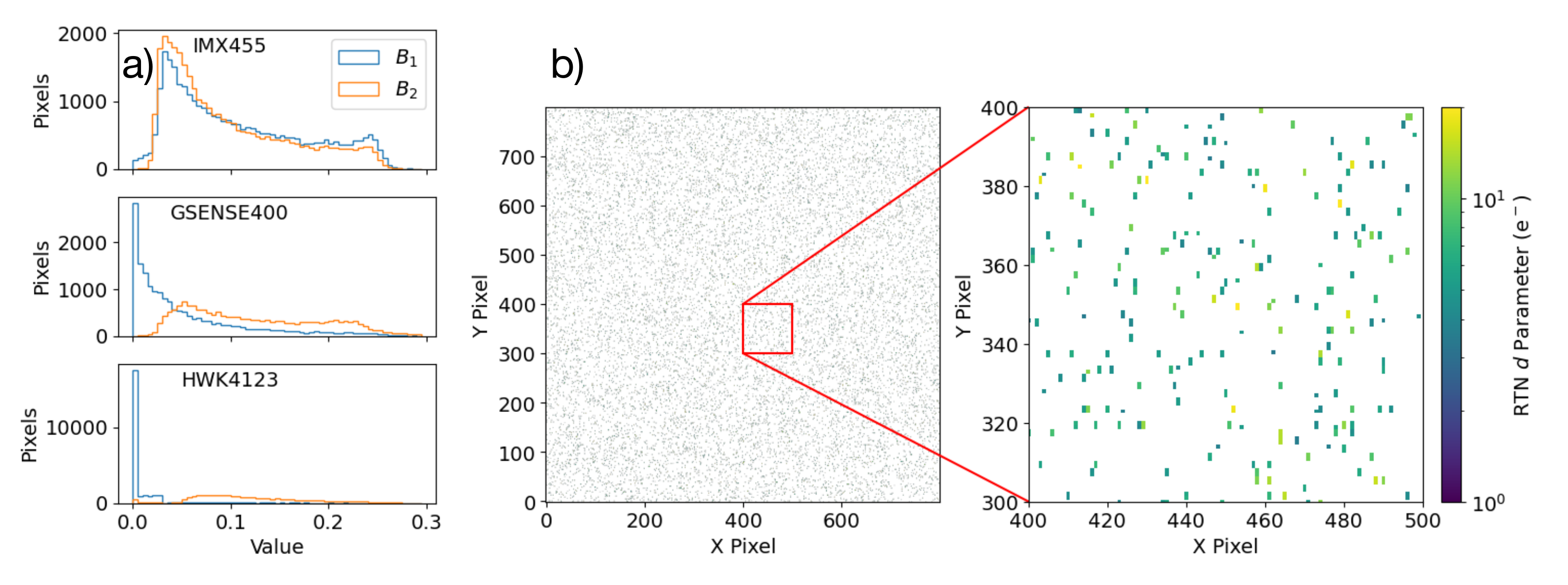}
    \caption{Distribution of pixels with random telegraph noise (RTN) across an 800$\times$800 subarray of camera QHY600M-C. Pixels with no RTN identified are white. The color of pixels with RTN indicates the best-fit value of the size of its RTN jumps, $d$. Pixels with RTN tend to be vertically adjacent to another pixel with RTN, with similar bias level distribution parameters $A, B_1, d, \sigma_r$}.
    \label{fig:rtn_spatial}
\end{figure}

In the 800$\times$800 pixel subarray analyzed within QHY600M-A, we identified 31980 pixels with correctable levels of RTN (5.0\% of the total number of pixels). This is a lower limit for the total number of pixels demonstrating RTN, due to the selection cuts described in Sec.~\ref{sec:app_fitting}. For example, pixels with two RTN-generating traps may not be well fit by a triple-Gaussian model. If $\sim$~5\% of pixels have a single trap, only $\sim 0.25\%$ of pixels are expected to have two traps. We identified slightly fewer RTN pixels for QHY600M-B and QHY600M-C: 24735 pixels (3.9\%) and 26984 pixels (4.2\%), respectively. The bias level distributions we observed in RTN pixels of the QHY600M cameras typically exhibited relatively high symmetry between the left and right Gaussian peaks, with the most common best-fit values of $B_1$ and $B_2$ both around 0.05 (see Fig.~\ref{fig:rtn_spatial}a, top panel).

We identified 17976 RTN pixels in the 800$\times$800 pixel subarray of the QHY42 (2.8\% of all pixels). Most of these pixels had asymmetric bias level distributions, with near-zero values of $B_1$ (see Fig.~\ref{fig:rtn_spatial}a, middle panel). Such suppression of low RTN jumps can occur if the period between correlated double samples is very short, as justified in \citet{wang:2006} by considering the evolution of the probability of trap occupancy (PTO). This suppression helps explain why RTN only causes a small increase in the RMS read noise in the QHY42: the read noise is only significantly increased by high RTN jumps. On the other hand, this behavior could lead to systematic overestimates of photon flux, as positive jumps are not at all balanced by negative jumps.

In the 800$\times$800 pixel subarray analyzed within the ORCA-Quest~2 camera, our RTN detection pipeline identified 24812 pixels (3.9\% of all pixels). To an even greater extent than the QHY42, the bias level distributions of these pixels had near-zero $B_1$ values, as shown in the bottom panel of Figure~\ref{fig:rtn_spatial}a. Moreover, the RTN jump magnitudes $d$ in the pixels of this camera are significantly smaller than the other cameras, with $d<1\,\mathrm{e}^-$ in almost all pixels. Thus, RTN in the HWK4123 sensor would not be effectively corrected by our algorithm, as the RTN jumps are smaller than even a single detected photoelectron. This feature likely arises from the ultra-low sense node capacitance designed into the HWK4123 sensor, which enables a single photoelectron collected at the sense node to create a larger voltage shift across the source follower transistor than a single electron captured or emitted in the oxide layer of this transistor.

The orange curves in Fig.~\ref{fig:read_noise_results} show the read noise distributions that the three cameras would have if the RTN that we identified was not present. Here, the original RMS read noise for each RTN pixel $\sigma_{r,tot}$ is replaced by its fit parameter $\sigma_r$. Qualitatively, we observe that RTN indeed contributes significantly to the extended tails in the read noise distributions. Quantitatively, we find that RTN increases the RMS read noise by $>19.6\%$ for the QHY600M-C camera, by $>4.2\%$ for the QHY42 camera, and by $>2.5\%$ for the ORCA-Quest~2 camera. Although the effects of RTN on the read noise distribution of the ORCA-Quest~2 camera are small, such subtle effects can be significant when attempting to perform photoelectron counting. When the read noise of a pixel is lower than $\sim 0.3\,\mathrm{e}^-$, one may begin to confidently resolve individual photoelectrons \citep{teranishi:2012}, thereby achieving an even lower effective read noise. In the case of the ORCA-Quest~2 camera, nearly all of the pixels with RTN (3.4\% of all pixels) would have a read noise below $0.3\,\mathrm{e}^-$ if only they did not have RTN. Therefore, while our algorithm may not be suitable for removing RTN from DSERN image sensors in post-processing, it is worthwhile for manufacturers of such sensors to seek to eliminate RTN in order to enable photon counting.

\subsubsection{Spatial distributions of RTN pixels}
\label{sec:spatial_distribution}
Figure~\ref{fig:rtn_spatial}b shows how RTN pixels are distributed across the subarray of QHY600M-C. Macroscopically, the RTN pixels appear randomly and isotropically distributed across the subarray. Thus, it could be challenging to position sources of interest to ensure they do not fall upon pixels with RTN, but would be possible for a single source as long as the PSF diameter is smaller than $\sim 8$~pix. Looking at a smaller subarray, as shown in Fig.~\ref{fig:rtn_spatial}, we also observe that it is highly likely for RTN pixels in the IMX455 to be vertically adjacent to a second RTN pixel with similar best-fit parameters $A, B_1, d, \sigma_r$. This pattern was also present in QHY600M-A and QHY600-B. This result might be explained if defects are more likely to be left behind during fabrication in pairs and if the fabrication occurs stepwise in the column direction. To determine whether there is any correlation between jumps in one RTN pixel and jumps in an adjacent RTN pixel, we measured the cross-correlation in bias signal levels for many pairs of such pixels. We found this cross-correlation to be consistent with zero for all pairs and for any frame offsets. Thus, the  RTN jumps themselves are localized to individual pixels. For the purposes of an RTN correction algorithm, therefore, the propensity for RTN pixels to come in vertical pairs is not relevant, but it merits future research to understand how to eliminate RTN during fabrication. The RTN pixels in the QHY42 did not exhibit this pairwise pattern and were randomly distributed across the subarray.

\subsection{Predicted benefit of RTN correction and pixel masking}
\label{sec:rtn_predicted_effects}
\begin{figure}
    \centering
    \includegraphics[width=0.8\linewidth]{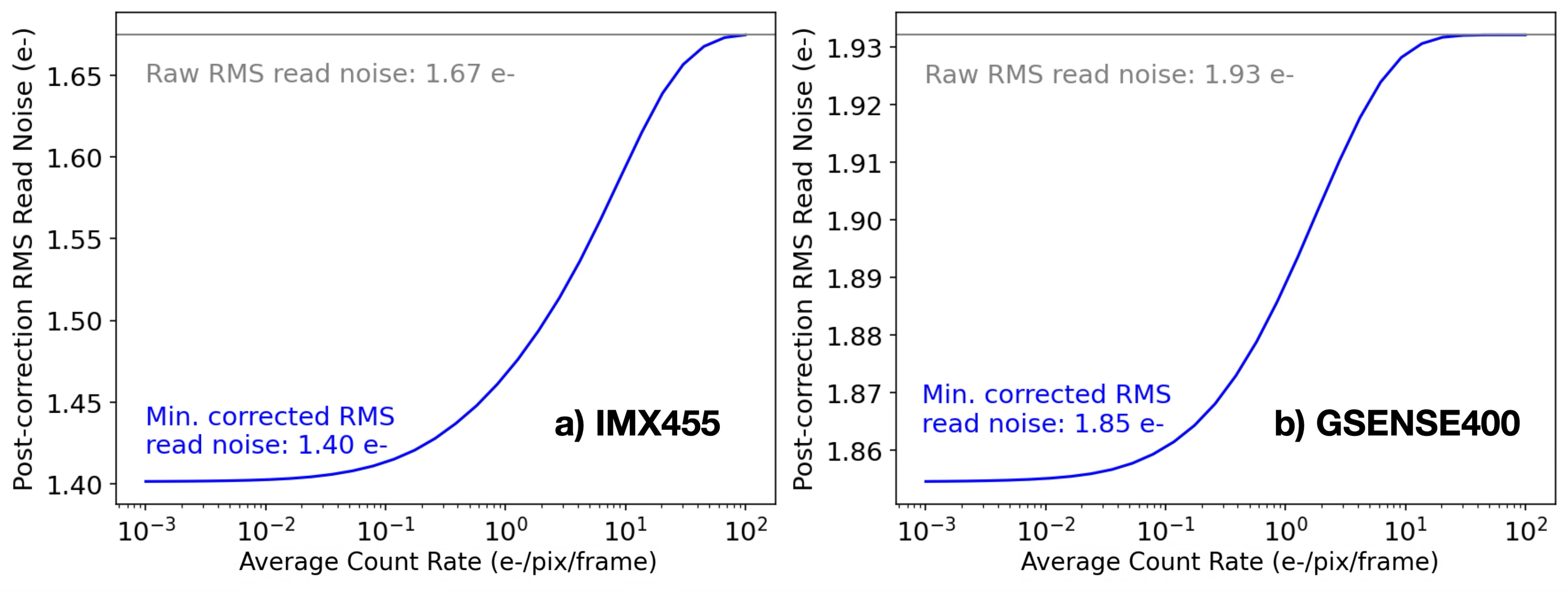}
    \caption{Predicted improvement deliverable by the RTN correction algorithm for \textbf{a)} the QHY600M-A camera housing an IMX455 sensor and \textbf{b)} the QHY42 camera housing a GSENSE400 sensor. The predicted improvement depends strongly on the average count rate in each pixel, as correction is not performed when Poisson noise becomes larger than the RTN jumps.}
    \label{fig:read_noise_improvements}
\end{figure}

As discussed in Sec.~\ref{sec:prediction_methods}, we predicted the improvement to sensor read noise that the RTN correction algorithm could deliver for each sensor as a function of the average electron count rate of each pixel. These predictions are shown in Fig.~\ref{fig:read_noise_improvements}a and Fig.~\ref{fig:read_noise_improvements}b for the QHY600M-A and QHY42 cameras, respectively. 

\begin{figure}
    \centering
    \includegraphics[width=0.9\linewidth]{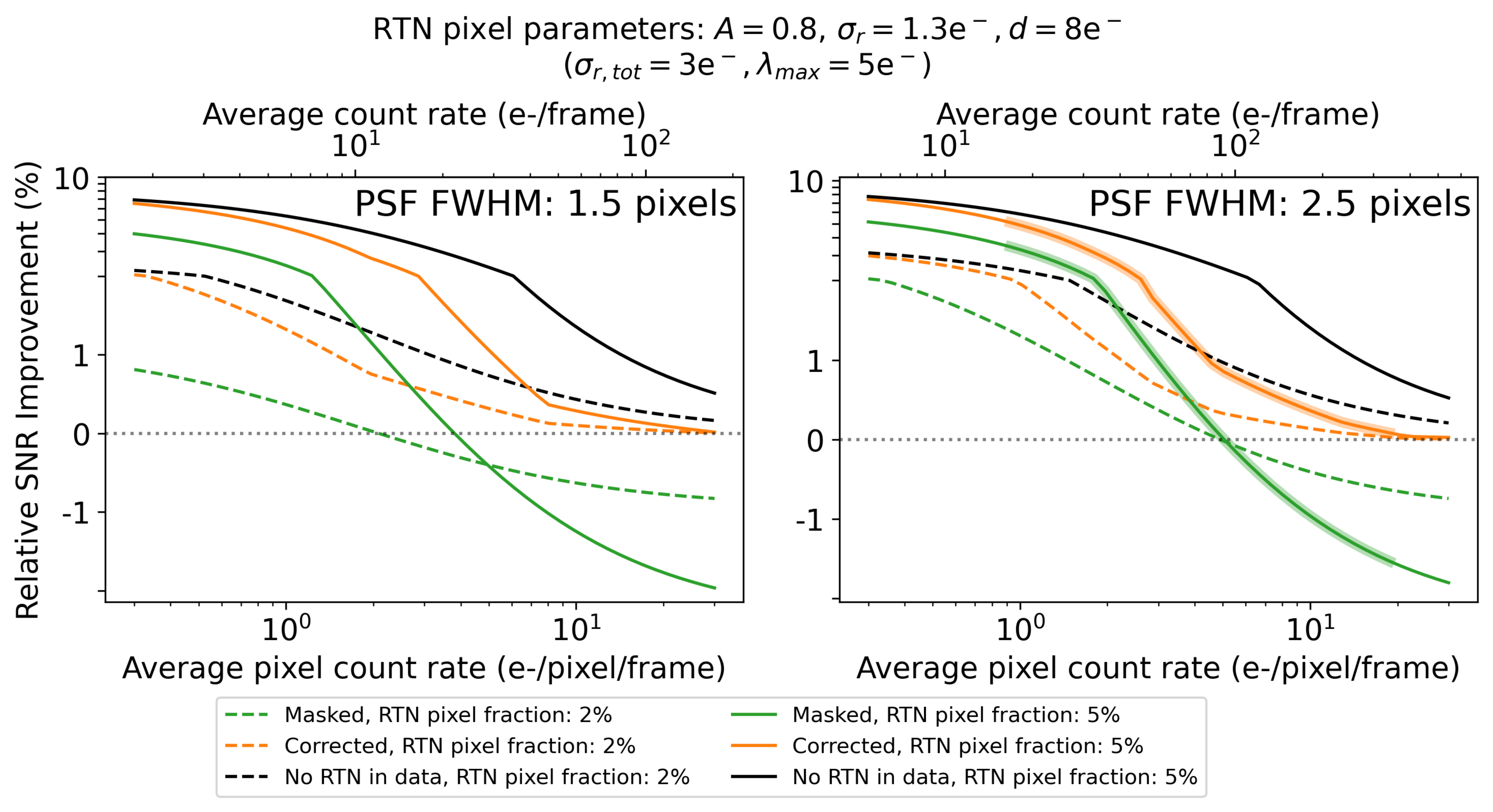}
    \caption{Predicted photometric improvements deliverable by the masking of RTN pixels (green curves), by our correction algorithm (orange curves), and by a theoretical perfect removal of RTN (black curves). All RTN pixels are assumed to have the same bias level distribution (here using average values for RTN pixels in the IMX455), with different line styles (solid, dashed) representing sensors with different fractions of pixels demonstrating RTN (5\%, and 2\%, respectively). In the left plot, an undersampled PSF is assumed; in the right, a critically sampled PSF. The highlighted segments in the right plot indicate the region probed with our on-sky tests using QHY600M-C, agreeing well with Fig.~\ref{fig:light_curve_improvements}b.}
    \label{fig:masking_comparison}
\end{figure}

We then performed Monte Carlo simulations to estimate how the RTN correction algorithm would affect the overall SNR of time-series observations of a single source. We simulated a Gaussian PSF falling upon a pixel array, where each pixel had some probability of having RTN, a large number of times. Using an aperture with radius equal to the PSF FWHM, we calculated the SNR of the source, accounting for shot noise and read noise in each pixel, first assuming no RTN correction was applied and then using the estimated post-correction read noise for RTN pixels. We also calculated the SNR that would be achieved if all RTN pixels were simply masked from the aperture as well as the SNR that would be achieved if no RTN was present. By taking the ratios of the masked, corrected, and no-RTN SNRs with the uncorrected SNR, we quantify the effectiveness of the RTN mitigation strategies relative to the theoretical maximum. The results are shown in Figure~\ref{fig:masking_comparison}, where we repeated this simulation for sources of varying brightness, for PSFs of varying size, and for sensors with varying fractions of RTN pixels.

We found that our correction algorithm (orange curves) outperforms masking (green curves) in all instances, although masking is nearly as effective when the PSF is well-sampled and read noise dominates the noise budget. When the PSF is undersampled or for brighter sources, the correction algorithm more dramatically outperforms pixel masking. Also, for many individual simulated sources, pixel masking actually degraded the SNR when the RTN pixels fell near the PSF core. This was observed in on-sky data as well---see Fig.~\ref{fig:light_curve_improvements}b. We therefore recommend our RTN correction algorithm for delivering consistent SNR improvements, especially when observing many sources across a wide range of brightnesses. For observations of one or a small number of sources, particularly in the read noise dominated regime, masking may provide adequate RTN mitigation. Alternatively, one might try to place the sources in avoidance of RTN pixels. RTN pixel identification is necessary for any of these mitigation options.

\subsection{RTN mitigation in real data}
\label{sec:on_sky}
We next applied our correction algorithm and pixel masking to real astronomical observations. For all of these applications of the algorithm, we used $\alpha=0.003$ and the temporal median technique, with a window size of $N=10$.

\subsubsection{Bias frames}
\label{sec:results_biases}
We first applied the algorithm to 1000 bias frames from QHY600M-A and the QHY42 that were not used in the RTN identification step. We compared the RMS read noise of the 1000 corrected bias frames to the predicted improvement. We found the RMS read noise of the corrected bias frames to be 1.43\,e$^-$ for the QHY600M-A and 1.84\,e$^-$ for the QHY42, in good agreement with our expected values of 1.40\,e$^-$ and 1.85\,e$^-$.

% The small discrepancy for QHY600M-A may be due to the simple method by which we predict the improvement delivered by the algorithm: even when there is no flux, some RTN jumps will go uncorrected for pixels that have RTN jump level $d$ just larger than $3\times \sigma_r$.

\subsubsection{Stellar light curves}
Next, using QHY600M-C mounted on an LCO small telescope, we observed a star field centered on $\alpha = 18^{\rm h}36^{\rm m}46^{\rm s}.08$, $\delta = -15\degr09\arcmin21\farcs6$. Using the same $800\times 800$~pixel subarray as used for RTN identification, we covered a field of size $9.85' \times 9.85'$. We imaged this field with an exposure time of 0.125~s for 4 minutes (2000 frames), then passed the exposures through the RTN correction algorithm. We used an aperture photometry pipeline based on the python package \texttt{photutils} to create original and corrected light curves for stars in this field. With this pipeline, we identified 766 stars in the field, measured their median full width at half maximum (FWHM) to be 2.6~pixels, and extracted their fluxes using an aperture with radius equal to the FWHM. We subtracted an average bias frame from both sets of frames. For corrected frames, we also subtracted bias correction values for pixels in the apertures of sources, as noted in see Sec.~\ref{sec:subtlety}. We employed median background subtraction using annuli with inner and outer diameters of $2.5\times$ and $4\times$ the FWHM. We also performed aperture photometry after masking all RTN pixels. On the time scale of $\sim 4$\,min, intrinsic stellar variability is minimal. Therefore, any variability in the light curves is attributed to Poisson noise, instrument noise, or atmospheric effects. We used the 20 brightest stars as comparison stars for all other stars in the field to mitigate atmospheric scintillation. To do so, we normalized the original light curves for these 20 stars, found the average normalized flux in each exposure, and divided all light curves (original, corrected, and masked) by these correction factors. This correction, when applied to the 20 comparison stars themselves, artificially removes $\approx 5\%$ of variability. As we are most interested in the improvements delivered by the RTN correction strategies, not the overall SNR, we still keep these comparison stars for further analysis.

\begin{figure}[ht!]
    \centering
    \includegraphics[width=0.99\linewidth]{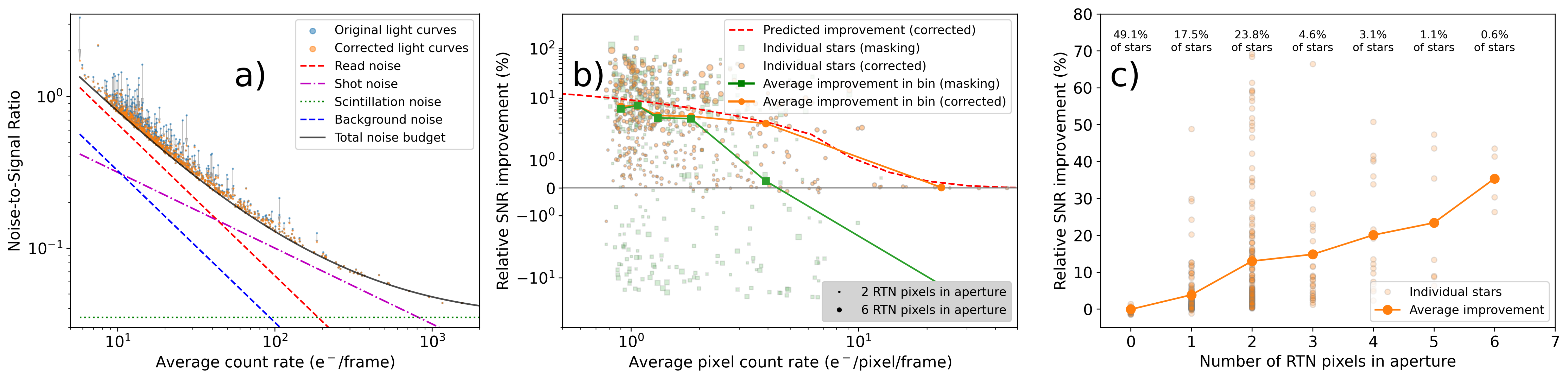}
    \caption{Photometric improvement delivered by the RTN correction algorithm for 766 stars observed with QHY600M-C. \textbf{a)} The noise-to-signal ratio (NSR) of each star's light curve before (blue points) and after (orange points) RTN correction vs. average stellar flux. Curves show the expected noise contribution from various sources. The correction moves most stars with higher-than-expected noise close to the expected total noise curve (black). \textbf{b)} Relative improvement to the SNR of each star delivered by RTN correction (orange circles) or RTN pixel masking (green squares). Marker sizes correspond to the number of RTN pixels in each aperture. Solid orange and green curves show the average improvements for bins of 150 stars, grouped by increasing flux. Stars with no RTN pixels in their aperture (and no SNR change) are not shown but are included when calculating averages. The red dashed curve shows the expected average improvement of the correction algorithm. \textbf{c)} Relative improvement to the SNR for faint stars ($< 3\,\mathrm{e}^-$/pixel/frame; orange circles) vs. the number of RTN pixels in the stars' apertures. The solid orange curve shows the average improvement for stars with a fixed number of RTN pixels in their apertures.}
    \label{fig:light_curve_improvements}
\end{figure}

We then calculated the SNR of the original, corrected, and masked light curves and found the improvements delivered by the correction and masking. Figure~\ref{fig:light_curve_improvements}a shows the noise-to-signal ratio (NSR; inverse of SNR) for the original and corrected light curves as a function of source brightness. For each star, a gray arrow connects the NSR of the original light curve (blue points) to the NSR of the corrected light curve (orange points). The curves show the expected noise contributions from known sources, with the total expected NSR (black curve) found by adding these sources in quadrature. The red dashed line represents the contribution from read noise, assuming an RMS value of $1.4\,\mathrm{e}^-$ (the value expected without RTN). Unlike the black points, nearly all of the green points lie on the blue curve, indicating that the algorithm removes most excess noise associated with RTN.

Figure~\ref{fig:light_curve_improvements}b illustrates the percent improvement in SNR that the correction algorithm (orange circles) and masking (green squares) provided for each star. The size of each circle/square is proportional to the number of RTN pixels in the star's aperture. We also calculated the average SNR improvements for sets of 150 stars, binned by increasing count rate (with the final bin containing only the brightest 16 stars). These average improvements are shown by the orange curve and green curve for the correction algorithm and masking, respectively. The orange curve agrees well with the algorithm's predicted SNR improvement (red dashed curve).

On average, both the correction algorithm and masking delivered SNR improvements of $> 5$\% for sources with an average pixel count rate of $< 3\,\mathrm{e}^-$/pix/frame. Larger improvements ($\gtrsim 10\%$) are expected for count rates $\lesssim 0.5\,\mathrm{e}^-$/pix/frame, but in our experiment the sky background itself contributed $\approx 0.5\,\mathrm{e}^-$/pix/frame. For individual stars, pixel masking showed higher variance than the correction algorithm, even degrading the SNR when masked pixels were near the PSF center. For brighter sources ($> 3\,\mathrm{e}^-$/pix/frame), masking consistently degraded the SNR (by 4\% on average), while the correction algorithm still delivered small SNR improvements (2\% on average).

The algorithm also yielded significantly larger SNR improvements for sources falling on a larger number of RTN pixels, as shown in Fig.~\ref{fig:light_curve_improvements}c. The relationship was similar for pixel masking.

As explained in Sec.~\ref{sec:subtlety}, using bias correction values for corrected images was necessary to avoid slightly overestimating the brightnesses of stars with RTN pixels. When we did not use these values, the mean flux of such stars was $0.4\,^{+0.9}_{-0.4}$\% greater in corrected images than original images (median, lower quartile, and upper quartile values). When we applied all appropriate steps, the correction algorithm maintained absolute photometry and boosted photometric precision. Masking consistently underestimates the brightness of stars with RTN pixels if no other corrections are used.

% As explained in Sec.~\ref{sec:subtlety}, if care is not taken, the average flux measured for each star can be artificially increased when using the correction algorithm. When we did not use bias correction values, we indeed observed a systematic increase in the measured flux of stars with at least 1 RTN pixel. The mean flux for such stars was $0.4\,^{+0.9}_{-0.4}$\% greater in corrected images than original images, where here we specify the median, lower quartile, and upper quartile. When, however, we used the correction values, the mean flux of these stars was the same ($-0.05\,^{+0.07}_{-0.17}$\% greater) in corrected and original images. Thus, when all appropriate steps are applied, the correction algorithm maintains absolute photometry and boosts photometric precision.
% XXX DRH: The contrast in figure 7 a is very low. hard to see what the improvement is. 
% CPL: Is it better now?

\subsubsection{Asteroid occultation light curve}
\label{sec:results_anchises}
Finally, on February 3, 2026, we observed the occultation of Trojan asteroid Anchises of a background star located at $\alpha = 01^{\rm h}27^{\rm m}02^{\rm s}.3092$, $\delta = +15\degr42\arcmin48\farcs837$ using QHY600M-A and QHY600M-B, mounted at McDonald Observatory. We observed the occulted star for $\sim 5$~min before and after the occultation, with exposure time of 0.1~s using QHY600M-A and 0.3~s using QHY600M-B. We made no attempt to position the PSF of the occulted star on pixels with or without RTN. We applied the RTN correction algorithm to both sets of frames, then used our aperture photometry pipeline to generate light curves for the occulted star, with five nearby stars serving as comparison stars.

\begin{figure}[ht!]
    \centering
    \includegraphics[width=0.9\linewidth]{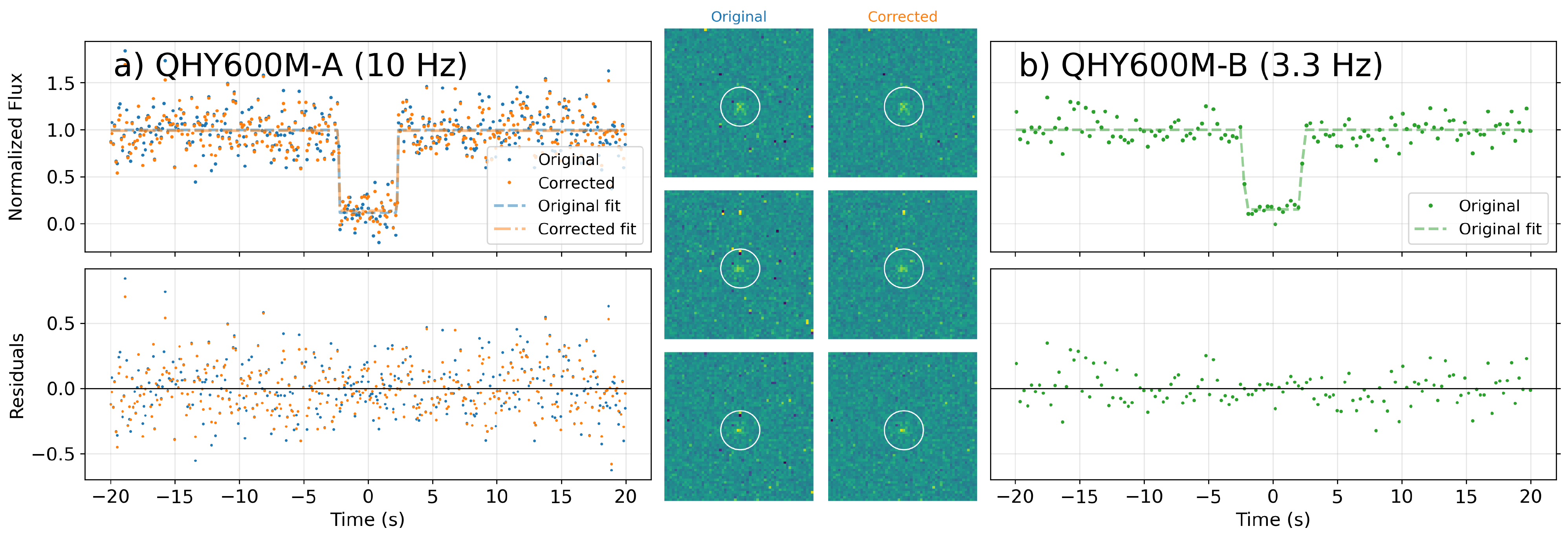}
    \caption{Light curves of a stellar occultation by Anchises, observed with two telescopes hosting QHY600M cameras. \textbf{a)} The occultation as observed by QHY600M-A with exposure time 0.1~s, both before (blue) and after (orange) RTN correction. Note that during occultation, more negative than positive RTN jumps occurred and were corrected. This directly affected the measured occultation depth. The middle panel contains three consecutive frames from a subarray around the target star before and after RTN correction, with the aperture shown as a white circle. Most salt-and-pepper noise in the original frames is removed in the corrected frames. \textbf{b)} The occultation as observed by QHY600M-B with exposure time 0.3~s, for which the target star aperture fell on no RTN pixels. Each light curve is fit to a square well with temporal smoothing. Bottom panels show residuals from these best-fit models.}
    \label{fig:anchises_occultations}
\end{figure}

For QHY600M-A, the aperture of the target star fell on four RTN pixels; for QHY600M-B, it fell on none. We therefore saw negligible differences between the original and corrected light curves for QHY600M-B. We plot the light curves of the target star for 40~s around the occultation in Fig.~\ref{fig:anchises_occultations}, only plotting the original light curve for QHY600M-B.

We fit a simple occultation model to these original and corrected light curves, consisting of a square well with temporal smoothing during ingress and egress, using the form
\begin{equation}
    F(t) = 1 - \frac{D}{2} \left[
    \mathrm{erf}\!\left(\frac{t - t_{\mathrm{in}}}{\tau}\right)
  - \mathrm{erf}\!\left(\frac{t - t_{\mathrm{eg}}}{\tau}\right)
\right]
    \label{eq:occultation_model}
\end{equation}
where $F(t)$ is the normalized flux of the star at time $t$, $D$ is the depth of the occultation, $t_{in}$ is the ingress time, $t_{eg}$ is the egress time, and $\tau$ is the smoothing time for the ingress and egress. This approximates a sharp occultation convolved with Fresnel diffraction, the finite projected stellar diameter, and the exposure time. A more thorough analysis would perform this convolution numerically \citep{widemann:2009}, but this approximation serves to demonstrate the performance of the correction algorithm. For each light curve, we calculate the RMS of the residuals for the best fit to this model.

Figure~\ref{fig:anchises_occultations} also shows the best-fit occultation models and the residuals to these fits (bottom panels). Table~\ref{tab:occultation_fits} shows best-fit parameters and RMS residuals for these models. The occultation depths are consistent with the expected brightness of Anchises. The algorithm delivered tighter errors on the fit parameters and an $\approx 7\%$ improvement to the RMS residuals. The improvement to the RMS residuals agrees well with the algorithm's predicted SNR improvement, given that the average pixel count rate outside of the occultation was $\approx3.5\,\mathrm{e}^-$/pix/s. RTN correction also caused the best-fit occultation model to have a noticeably shallower depth, in better agreement with the depth found using QHY600M-B. Upon inspection of the frames during occultation, it was apparent that there were by coincidence more negative than positive RTN jumps during the occultation, which resulted in the original light curve delivering an overestimate of the occultation depth. This occultation observation provides further proof that the correction algorithm can improve the data quality of astronomical instruments hosting detectors with RTN.

\begin{table}[ht]
\centering
\caption{Fit results for the occultation of Anchises observed with QHY600M-A and QHY600M-B.}
\label{tab:occultation_fits}
\begin{tabular}{lcccc}
\hline\hline
Dataset & Duration (s) & Depth (\%) & Smoothing time $\tau$ (s) & RMS Residuals \\
\hline
0.1s Original (QHY600M-A)       & $4.55 \pm 0.04$ & $88.3 \pm 3.3$ & $0.01^{+0.01}_{-0.01}$ & 0.207 \\
0.1s Corrected (QHY600M-A) & $4.55 \pm 0.04$ & $86.0 \pm 3.0$ & $0.02^{+0.01}_{-0.02}$ & 0.192 \\
0.3s Original (QHY600M-B)       & $4.50 \pm 0.03$ & $84.7 \pm 3.2$ & $0.02^{+0.01}_{-0.01}$ & 0.120 \\
\hline
\end{tabular}
\end{table}

\section{Algorithm Usage}
\label{sec:usage}
The software that we used to identify RTN pixels, fit the bias distributions of these pixels, and apply the correction algorithm is publicly available in the GitHub repository \texttt{\href{https://github.com/ChrisLayden/RTN-corrector}{RTN-Corrector}}. One python script, \texttt{rtn\_fitter}, performs RTN identification and parametrization using a set of bias frames collected by the user. This script creates a FITS file containing RTN fit parameters as well as representative plots demonstrating the extent to which RTN contributes to the noise floor of the sensor. Figures~\ref{fig:read_noise_results} and \ref{fig:read_noise_improvements} are examples of some of these plots. The gain of the camera, in ADU/e$^-$, must be specified in order for \texttt{rtn\_fitter} to save pixel bias level distributions and to predict the noise improvements possible at varying levels of illumination. If the gain is not specified, \texttt{rtn\_fitter} will report the distribution of pixel read noise values (in ADU) and the number of RTN pixels identified. A second python script, \texttt{rtn\_fixer}, applies the RTN correction algorithm to a stack of science frames, using the pre-computed RTN fit parameters. Science frames and bias frames should use the same gain setting, readout mode, and operating temperature.

The settings that we have used in this work and recommend for most circumstances are $\alpha = 0.003$ (for which we have pre-computed a LUT), $N=10$, and the ``temporal median" technique for generating reference values. With these settings, the algorithm attempts to correct jumps in RTN pixels with $d/\sigma_r>3$, as long as $\lambda <\lambda_{max}$ for the pixel and the variability in the source over $N$ frames is $\lesssim \sqrt{\lambda}$.

\subsection{Data volume and speed considerations}
\label{sec:data_volume}
As discussed in Sec.~\ref{sec:identification}, 1,000--5,000 bias frames are recommended to identify correctable RTN pixels and robustly parametrize their bias distributions. For large-format cameras or instruments using many cameras, this can be an unwieldy amount of data. If the bias data set is too large relative to the user's available random-access memory, \texttt{rtn\_fitter} can split the sensor area into slices, loading and fitting data from each slice one at a time. If simply storing this data volume is untenable, the user might first collect just 400 bias frames to identify the subset of pixels with non-normal bias level distributions (see Sec.~\ref{sec:app_fitting}). The user could then collect the full set of bias frames, keeping only the readings of non-normal pixels. This can reduce the requisite data volume by over an order of magnitude, depending on the sensor.

% For space observatories, pixels may develop RTN over time due to radiation damage. In this case, one might perform RTN fitting at fixed intervals of operation, requiring minimal observing time (bias frames could even be collected during slewing). Data downlink concerns could be mitigated because bias frames are inherently highly compressible and because one might save the full set of bias frames only for pixels identified as having non-normal bias level distributions, as discussed above.

As RTN mitigation is especially relevant for high-speed imaging, the correction algorithm should not bottleneck data processing. On a moderately powerful desktop computer, with the settings used in our analysis, \texttt{rtn\_fixer} processed science frames for the $800\times 800$ subarray of QHY600M-A at a rate of $\sim 77$~FPS. As this would correspond to a rate of $\sim 0.75$~FPS for the full 66 megapixel sensor, we have developed an alternative script for users of large-format cameras that is optimized for graphics processing unit (GPU) computing. This script, \texttt{rtn\_fixer\_gpu}, processed the $800\times 800$ frames at $\sim 200$~FPS, corresponding to $\sim 2$~FPS for a full IMX455 sensor. If greater speed is needed, minor changes could be made to apply the correction algorithm only to RTN pixels in pre-identified sources, instead of to all RTN pixels in the original frames.

\section{Conclusion}
We have developed and implemented a Python-based algorithm to mitigate RTN (also known as ``salt-and-pepper" noise) in solid state detectors that use correlated double sampling. The mitigation is based on \textit{a priori} per-pixel characterization of the RTN noise. The algorithm is most useful when the exposure level is lower than a pixel's RTN jump, such as in high speed data acquisition or low flux/background situations (e.g., ultraviolet imaging). Applications include improvement of time series photometry and removal of salt-and-pepper noise in the backgrounds of images. We have also shown that simply masking RTN pixels can deliver similar photometric improvements in some regimes. Our open-source software for identifying and parametrizing RTN in sensor pixels using a stack of bias frames is useful for both pixel masking and our correction algorithm.

Our de-noising algorithm may also be applicable to other scientific domains involving time-series imaging, as it is straightforward, efficient, independent of the density of RTN pixels, and retains high fidelity of source brightness and morphology. Other de-noising techniques frequently used in image processing may be computationally cumbersome or may introduce blur or distortion \citep{chen:2020}.

We have found that in three commonly used CMOS image sensors---the Sony IMX455, the Gpixel GSENSE400, and the Fairchild Imaging HWK4123---RTN increases the read noise floor by $\gtrsim 20\%$, $\gtrsim 4\%$, and $\gtrsim 2.5\%$, respectively. Although we identified RTN in $\sim 3\textrm{--}5\%$ of pixels for all three cameras, the bias level distributions of RTN pixels in the latter two cameras were more one-sided and compact. These facts, likely caused by readout timing and source follower transistor design, explain the difference in the overall effect of RTN between these sensors and the IMX455.

We confirmed that our RTN correction algorithm performs as expected by applying it to real astronomical observations taken with the LCO small telescope network, which employs IMX455 sensors. For observations of faint sources ($< 3\,\mathrm{e}^-$/pix/frame) the algorithm delivered an average SNR improvement of $> \,5\%$. Pixel masking provided similar improvements on average but occasionally degraded photometry. RTN mitigation may be even more effective for sensors in space telescopes, which may develop an increased number of RTN pixels over time due to radiation damage.

\section*{Disclosures}
We declare no financial, commercial, or other conflicts of interest related to the research presented in this paper.

\section*{Code, Data, and Materials Availability}
The software developed and used in this work is available in the \href{https://github.com/ChrisLayden/RTN-corrector}{RTN-Corrector} GitHub repository. This repository also contains a representative sample of the bias frames obtained from the QHY600M-C camera used in this work (5000 frames of a $100\times 100$~pix subarray for each camera). It also contains a representative sample of on-sky data obtained with the QHY600M-C camera (1000 low-illumination frames for the same $100\times 100$~pix subarray). The remainder of the data generated and analyzed in this study are not publicly available due to their large volume. However, this data is available from the corresponding author upon reasonable request.

\section*{Acknowledgments}
This research made use of Photutils, an Astropy package for
detection and photometry of astronomical sources \citep{bradley:2024}. This research made use of the Lucky Star Project \href{https://lesia.obspm.fr/lucky-star/}{occultation predictor}, which uses ephemerides for various outer solar system objects from \cite{desmars:2015}. This paper is based on observations made with the Las Cumbres Observatory’s education network telescopes that were upgraded through generous support from the Gordon and Betty Moore Foundation. We made use of the large language model-based agentic coding tool Claude Code to improve the performance of software developed in this paper. All changes recommended by agentic coding tools were inspected and validated by the authors.

\bibliography{references}{}
\bibliographystyle{aasjournal}

\appendix
\section{RTN Pixel Identification and Parameter Estimation}
\label{sec:app_fitting}
\begin{figure}[h!]
    \centering
    \includegraphics[width=0.5\linewidth]{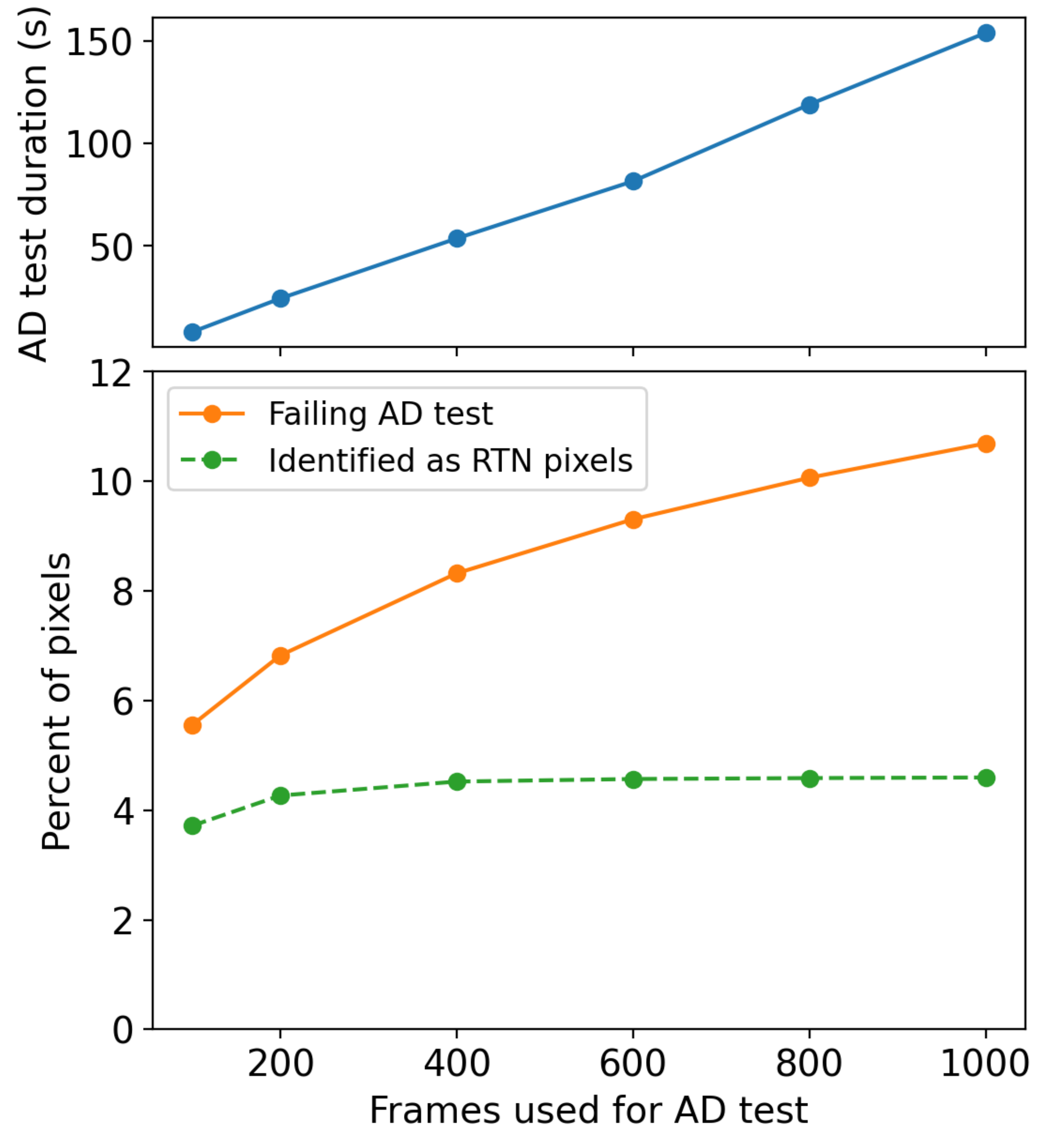}
    \caption{Effect of increasing the number of frames used for AD normality testing on the time required for the test (top), the fraction of pixels in the array failing the AD test (bottom, orange), and the fraction of pixels identified as exhibiting RTN after failing the AD test and undergoing triple-Gaussian fitting (in all cases, 5000 bias frames were used for fitting). Results were found using QHY600M-C.}
    \label{fig:ad_scaling}
\end{figure}

Here we outline the steps performed by \texttt{rtn\_fitter} to
identify and parametrize RTN in individual pixels using a stack of many bias frames. We recommend using between 1000 and 5000 bias frames (this recommendation is justified below). We first apply the Anderson-Darling (AD) normality test to each pixel's bias readings, following \citet{ozdogru:2025}. We keep any pixels with an AD statistic above 1.092 \citep[the 99\% confidence limit for identifying non-normality in a sample with unknown population mean and variance;][]{stephens:1974} for further testing of RTN. Again following \citet{ozdogru:2025}, we apply uniform smoothing to the digital bias values before this test. Regardless of the total number of bias frames available, we apply the AD test to at most the first 400 frames. This cutoff prevents the AD testing from taking long relative to the triple-Gaussian fitting step, as the duration of the AD test scales linearly with the number of frames (see Fig.~\ref{fig:ad_scaling}). Also, as the number of frames used increases, more pixels without RTN but with slightly non-Gaussian behavior (including one-off outliers potentially from cosmic rays) fail the AD normality test, unnecessarily increasing the time required for RTN fitting. As shown in Fig.~\ref{fig:ad_scaling}, when more than 400 frames are used for the AD test, more pixels are retained for fitting, but a negligible number of these pixels are then identified as having RTN.

For every pixel failing the AD normality test, we create a normalized histogram of the pixel bias level readings, using the full set of 1000--5000 frames. To prevent the digital nature of these readings from affecting the histogram, the bin widths are set to integer values spanning the minimum and maximum reading for each pixel. As close to 30 bins as possible are used while maintaining integer bin widths. For the ORCA-Quest~2 camera, due to a quirk during digitization, some digital values are never reported in each pixel \citep{gallagher:2024}. Thus, some bins in the bias level histograms for pixels in this camera were empty, despite neighboring bins being non-empty. As these gaps negatively affected fitting performance, for the ORCA-Quest~2 only, we replaced the zeroes such bins with the average value of the two neighboring bins.

We use the \texttt{scipy.curve\_fit} library to attempt to fit each histogram to a triple-Gaussian distribution. This distribution is defined as
\begin{equation}
    f(x) = A\mathcal{N}(\mu_b, \sigma_r) + B_1\mathcal{N}(\mu_b - d, \sigma_r) + B_2 \mathcal{N}(\mu_b+d, \sigma_r),
\label{eq:triple_gaussian}
\end{equation}
and can be visualized in Fig.~\ref{fig:rtn_hist}. This yields best-fit values and standard errors for $A, B_1, B_2, \mu_b, d,$ and $\sigma_r$. We normalize the best-fit values and standard errors for $A, B_1$, and $B_2$ by dividing each by $(A+B_1+B_2)$. We then determine that a pixel has correctable RTN if its fit was successful and satisfied the following conditions:

\begin{itemize}
    \item $A>B_1$ and $A>B_2$,
    \item $A<0.95$,
    \item $d \geq 3\sigma_r$,
    \item The standard errors for all fit parameters (except $\mu_b$) are no more than $1/3$ of the corresponding best-fit values. If $B_1 < 0.03$, the standard error for $B_1$ need not satisfy this condition.
\end{itemize}

The first criterion ensures the fit correctly identified the central Gaussian. The second ensures that the side peaks correspond to RTN jumps that occur at a statistically significant frequency, not random outliers like cosmic ray hits. The third ensures that the pixel has a level of RTN that is correctable, with $3\sigma_r$ corresponding to the $\alpha=0.003$ significance level discussed in Sec.~\ref{sec:jump_id}. The final condition ensures the fit was robust. The fit parameter $\mu_b$ is excepted from this relative quality cut because the value of $\mu_b$ is set by the black offset, which is not physically meaningful. The $B_1$ exception accounts for bias level distributions with single-sided RTN. Such distributions, where $B_1$ is negligible while $B_2$ is significant, can occur if there is a very short time interval between CDS samples \citep{wang:2006}. For all pixels identified as having correctable RTN, we saved the fit values $A, B_1, \mu_b, d,$ and $\sigma_r$ after converting $d$ and $\sigma_r$ to units of electrons by dividing by the sensor gain $K$ in ADU$/\mathrm{e}^-$.

\begin{figure}
    \centering
    \includegraphics[width=0.8\linewidth]{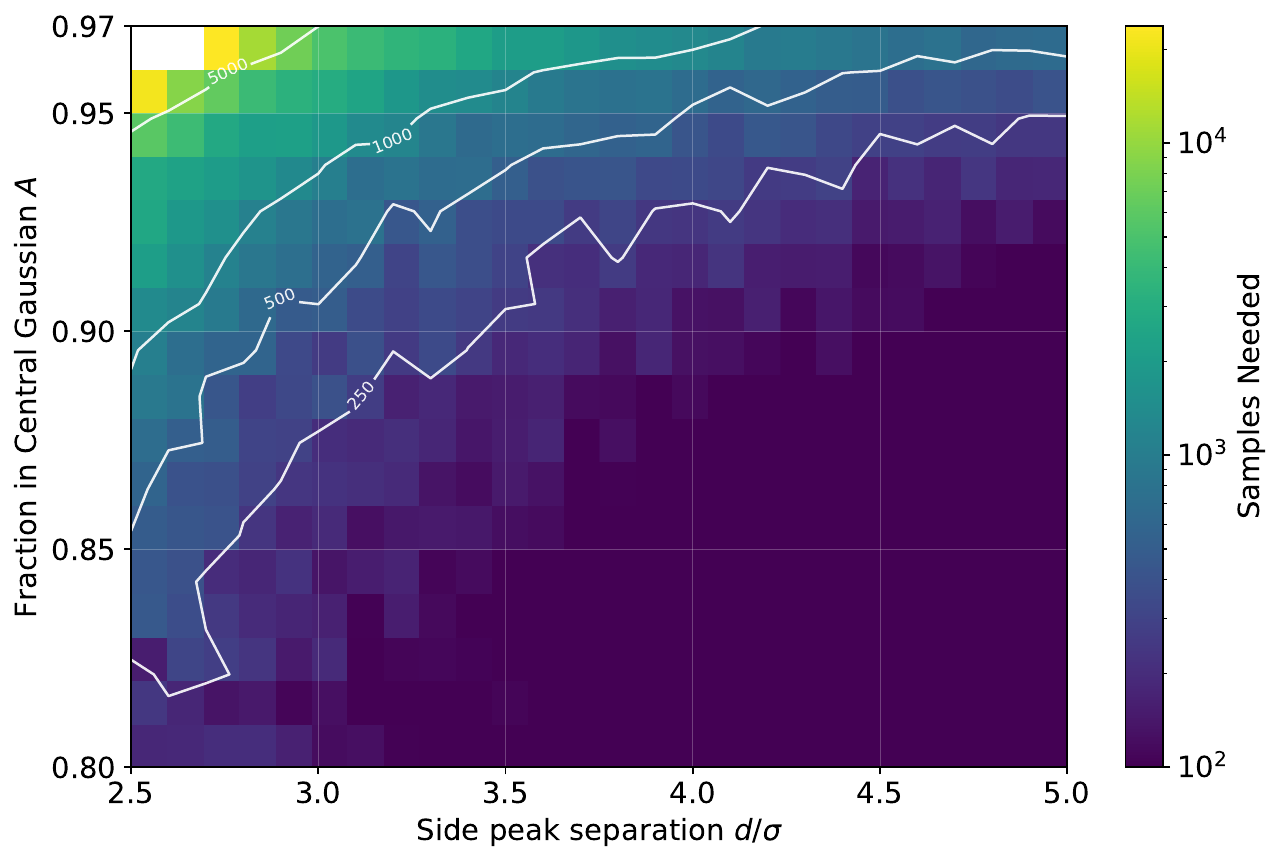}
    \caption{Number of samples needed to robustly fit simulated pixel bias level distributions with certain $A$ and $d/\sigma$ values. For this simulated data, a fit is deemed robust if the fit parameters have a standard error of at most 1/5 of the corresponding best-fit values.}
    \label{fig:sampsneeded}
\end{figure}

To understand how many bias frames should be used in RTN fitting, we performed Monte Carlo simulations in which we generated mock bias readings for pixels with varying RTN parameters. We varied parameters $A$ between 0.8 and 0.97 and $d$ between 2.5 and 5 times $\sigma_r$ (which was set to 1~e$^-$). We considered only symmetric bias level distributions, with $B_1=B_2=(1-A)/2$. For each pair of $A, d/\sigma_r$, we generated mock bias readings with size $N$ ranging from 100 to 10,000. We then determined the size $N$ at which the standard errors for all fit parameters were no more than 1/5 the corresponding best-fit values. This cutoff was chosen to ensure that for real data with additional complexity, fitting would satisfy the criteria above. Figure~\ref{fig:sampsneeded} shows the number of samples $N$ required for adequate fitting of pixels with varying RTN parameters. From this figure, we see that at least $N\gtrsim1000$ bias frames are needed to reliably parametrize RTN in pixels with $A \leq 0.95$ and $d/\sigma_r \geq 3$. We recommend (and have in this work used) a more conservative value of $N=5000$ to ensure that all RTN pixels in this regime are identified.

\end{document}